\documentclass[12pt]{article}
\usepackage{graphicx}
\newcommand{\beq}{\begin{equation}}
\newcommand{\eeq}{\end{equation}}
\newcommand{\bea}{\begin{eqnarray}}
\newcommand{\eea}{\end{eqnarray}}

\def\sq{{\vbox {\hrule height 0.6pt\hbox{\vrule width 0.6pt\hskip 3pt
   \vbox{\vskip 6pt}\hskip 3pt \vrule width 0.6pt}\hrule height 0.6pt}}}

\begin{document}
\begin{titlepage}

\begin{flushright}
{\tt hep-ph/0510334}\\
FIT HE-05-04 \\
RIKEN-TH-55 \\
SAGA-HE-222 \\
\end{flushright}
\vspace*{3mm}
\begin{center}
{\bf\Large Holographic Model for Hadrons \\
\vspace*{3mm}
in Deformed AdS${}_5$ Background \\ }
\vspace*{12mm}
Kazuo Ghoroku\footnote[1]{\tt gouroku@dontaku.fit.ac.jp},
Nobuhito Maru\footnote[2]{\tt maru@riken.jp, 
{\rm Special postdoctoral researcher, \\
\hspace*{7mm}Address after November 1st:
Dipartimento di Fisica, Universit\`a di Roma 'La Sapienza', \\
\hspace*{7mm}P.le Aldo Moro~2, I-00185 Rome, Italy}}, 
Motoi Tachibana\footnote[3]{\tt motoi@cc.saga-u.ac.jp},
Masanobu Yahiro\footnote[4]{\tt yahiro2scp@mbox.nc.kyushu-u.ac.jp}
\vspace*{2mm}

\vspace*{7mm}

{${}^{*}$Fukuoka Institute of Technology, 
Fukuoka 811-0295, Japan} \\
\vspace*{2mm}
{
${}^{\dagger}$Theoretical Physics Laboratory, RIKEN, 
Saitama 351-0198, Japan} \\
\vspace*{2mm}
{
${}^{\ddagger}$Department of Physics, Saga University, 
Saga 840-8502, Japan} \\
\vspace*{2mm}
{
${}^{\S}$Department of Physics, Kyushu University, 
Fukuoka 812-8581, Japan} 
\end{center}

\vspace*{20mm}

\begin{abstract}
 Several physical quantities of light hadrons are examined by a new
holographic model of QCD, which is the modified version of
the one proposed by Erlich et al. defined on AdS${}_5$. 
In our model, AdS${}_5$ is deformed
by a non-trivial bulk scalar, this is corresponding to adding the mass term of the
adjoint fermions to the 4d SYM theory dual to the gravity on 
AdS${}_5$. We find that this deformation should be taken to be rather 
small, but its important effects are also seen.
\end{abstract}
\end{titlepage}

\section{Introduction}
Recently, the gravity/gauge correspondence has revived the expectation
that QCD can be described by a string theory with an appropriate 
combination of D-branes. Some models are shown in
the system of $D_p/D_{p+4}$ branes \cite{KK}-\cite{GY2}. In these, by
setting the 10d background with stacked $D_p$ branes, $D_{p+4}$ branes
are embedded as a probe
to introduce flavor quarks. Then several physical quantities
of QCD have been obtained with sufficient values.

\vspace{.3cm}
When these models are set as 5d theories, we should prepare two kinds of
5d actions, $S_{\rm bulk}$ and $S_{\rm meson}$. 
The former describes the bulk background or closed
strings and the latter the hadrons or the system of open strings. 
They are corresponding to 10d action of gravity
and the one of $D_{p+4}$ branes, respectively. The solution of $S_{\rm bulk}$
gives the gravity-background dual to the Yang-Mills theory. 
On the other hand,
the classical solution of $S_{\rm meson}$ provides parameters
related to quarks, and the meson spectra are obtained from 
fluctuations around this solution in $S_{\rm meson}$.
However, there is no satisfactory holographic model, which describes QCD,
at present. Instead, several authors have proposed
phenomenological 5d-models which 
explain a wide range of hadron properties related to QCD \cite{EKSS,RP,TB,HS}.
In these models, the gravity-background is taken as AdS${}_5$ 
and this space is cut off at an appropriate infrared (IR) point in
order to realize the quark-confinement of the dual gauge theory.
However, 
true gravity-background
dual to QCD should be largely deformed from AdS${}_5$ in the infrared region
since the AdS${}_5$ background is dual to ${\cal N}=4$ 
supersymmetric Yang-Mills (SYM) theory.
Therefore,
even if the infrared region is removed by the IR cutoff mentioned above, 
we expect that the deformation 
of the background configuration from AdS${}_5$ is observed and it
would affect the physical quantities. 

Our purpose is to make clear this point in the approaches mentioned above.
This would be important to proceed these approaches to the next step.

\vspace{.3cm}
Here $S_{\rm meson}$ is set as the form used in \cite{EKSS}.
In this action, the mass and the vacuum expectation value (VEV) of bilinear
fields of quarks are given through the classical solution of a tachyonic scalar field. 
As mentioned above, we deform AdS${}_5$ 
by introducing the mass of adjoint fermions into the dual gauge theory.
This is realized 
by adding a non-trivial scalar, whose conformal dimension ($\Delta$) is three, 
to $S_{\rm bulk}$. As a result, in the dual gauge theory,
the supersymmetry is explicitly
broken and the gauge coupling constant is
running. So a new parameter is introduced in our model
as the mass of the adjoint 
fermions, through which we examine its dynamical effects on various
physical quantities related to light mesons.

\vspace{.3cm}
In the next section, we give our bulk-background configuration.
In section 3, $S_{\rm meson}$ and the configuration of
the scalar of $\Delta=3$ are given, and physical quantities of 
light mesons are examined to see effects of our deformation.
In the final section, summary is given.

\section{Bulk action and deformed SYM}
Up to now, many interesting 5d supergravity solutions have
been studied as deformed SYM theories. 
Here we adopt a simple but nontrivial model given in \cite{GTU},
which is briefly reviewed. 
Consider the following 5d action with a scalar ($\phi$),
\bea
   S_{\rm bulk}=\int d^4\!xdz\sqrt{-g}
   \left\{{1\over 2\kappa^2}(R-2\Lambda) 
    -{1\over 2}(\partial\phi)^2-V(\phi)\right\}  \ ,
                                                     \label{acg}
\eea
\beq
     V(\phi)=-{9\over 2}{\mu^2\over \kappa^2}
               \sinh^2\left\{\sqrt{{\kappa^2\over 3}}\phi\right\}
               \ .            \label{scalarp2}
\eeq
where $\mu=\sqrt{-{\Lambda}/6}$. The parameters
$\kappa^2$ and $\Lambda$ denote the five-dimensional gravitational 
constant and the cosmological constant, respectively.
From (\ref{scalarp2}), we can see that the mass of 
$\phi$ is $M^2=-3\mu^2$, then it corresponds to the 
conformal dimension three ($\Delta=3$) operator of
${\cal N}=4$ SYM theory \cite{gppz1}.

We can solve equations of motion for metric and $\phi$
under the ansatz, $\phi=\phi(z)$ and
\bea
   ds^2=g_{MN}dx^M dx^N=A^2(z)\left(\eta_{\mu\nu}dx^{\mu}dx^{\nu}
         +dz^2\right) \ .   
\label{fmet}
\eea
Here $\{M,N\}$ and $\{\mu,\nu\}$ denote the 5d and 4d indices,
respectively, and our metric convention is $\eta_{MN}=(-++++)$.
Setting as $$\lambda={\kappa^2\alpha^2\over 3},$$ where 
$\alpha$ is a constant, we obtain the following solution
\beq
A(z) = {\sqrt{\lambda}\over 
   \mu~\sinh\left[\sqrt{\lambda}z\right] }, \quad 
\phi(z) = \alpha z.
\label{scalar-bulk}
\eeq
From the second equation of (\ref{scalar-bulk}), we can see that
the parameter $\alpha$ represents the mass of the adjoint
fermions of ${\cal N}=4$ SYM theory \cite{gppz1}
since $\phi$ corresponds to the operator of $\Delta=3$ as mentioned above.
In this sense, the supersymmetry
of the gauge theory is broken for non-zero $\alpha$, and the gauge coupling would
be running.

\section{Meson spectra}
To discuss the meson properties, we start with
the 5d action of the theory with the background 
(\ref{fmet}) and (\ref{scalar-bulk}).
The fields in the bulk we consider here are the gauge fields, 
$L_M$ and $R_M$, and a scalar field $\Phi$ whose VEV is connected to 
chiral symmetry breaking.
In this letter, we focus on $N_f =2$ flavors. Then $\Phi$ transforms 
as a $({\bf 2_L}, {\bf 2_R})$. The action is
\begin{equation}
S_{\rm meson}=\int d^4x dz\, \sqrt{-g}\, {\rm Tr}\left[-\frac{1}{4g_5^2} ({L_{MN}L^{MN}}
+{R_{MN}R^{MN}}) - {|D_M\Phi|^2} - M^2_\Phi|\Phi|^2\right]\, ,
\label{3-1}
\end{equation}
where the covariant derivative is defined as 
$D_M\Phi=\partial_M \Phi+iL_M\Phi-i\Phi R_M\, $ and
$g$ is the determinant of the metric. $L_M=L_M^a\tau^a$, 
where $\tau^a$ are the Pauli matrices and
similarly for other fields. $g_5$ is the 5d gauge coupling.
We define $\Phi=S\, e^{i\pi^a\tau^a}$ 
and ${1\over 2}v(z)\equiv \langle S \rangle$, where
 $S$ corresponds to a real scalar and $\pi$ to a real pseudoscalar
($S\rightarrow S$ and $\pi\rightarrow -\pi$ under  $L\leftrightarrow R$).
They transform  as  ${\bf 1+3}$ under SU(2)$_V$.

\subsection{Scalar field}
\noindent {\bf chiral symmetry breaking:}~
Firstly let us study $v(z)$ in the case of our background. 
Here and hereafter we take $\mu =1$. We put
$M_{\Phi}^2 =-3$ following \cite{EKSS, RP}, which corresponds in the
CFT side to an operator with the conformal dimension $\Delta=3$, 
through the general relation $\Delta = 2+\sqrt{4+M_{\Phi}^2}$.
Solving the bulk equation of
motion for $S$,
\beq
   ({\sq}_5-M_{\Phi}^2)S=0  \,, 
   \label{mscalar}
\eeq
where $\sq_5$ denotes the five dimensional Laplacian, we obtain \cite{GY}
\beq
 v(z)={1\over A^3(z)}\left(c+m_q\left[-\lambda\sinh^{-1}(A(z)/\sqrt{\lambda})
       +{A(z)}\sqrt{A^2(z)+\lambda}\right]\right).
\label{3-2}
\eeq
Here $m_q$ and $c$ should be identified with the quark mass (explicit
breaking of the chiral symmetry) and the chiral condensate 
(spontaneous breaking of chiral symmetry), respectively. 
Actually this is approximated as $v(z)\sim m_qz+cz^3$ near $z=0$,
which reproduces the results of \cite{EKSS, RP}. We notice
that $v(z)$ includes the parameter $\lambda$ which is specific to
our model as well as $m_q$ and $c$. In other words, the gauge theory 
considered here is characterized by these three parameters.


\vspace{.3cm}
\noindent {\bf $\sigma$-meson:}~ As for the singlet meson state ($\sigma$), 
it is obtained as a solution for the fluctuation of $S$, $S=v(z)/2+\sigma$,
with finite four dimensional mass $m^2$ defined as 
$\partial_{\mu}\partial^{\mu}\sigma=m^2\sigma$. The equation for $\sigma$
is explicitly given in the same form with Eq.~(\ref{mscalar}) as
\bea
\frac{1}{A^2}\left[ m^2 
+ \partial_z^2 + 3\frac{\partial_z A}{A} 
\partial_z \right]\sigma = M^2_{\Phi} \sigma. 
\label{scalareom}
\eea
\\
It should be noticed that
this equation is independent of $v(z)$ and depend on $M^2_{\Phi}$ 
and $\lambda$ through the warp factor $A$.

Also in this case, the bound state spectrum
is obtained by introducing an IR cut-off, $z_m$,
into the fifth dimension \cite{EKSS,RP}, which corresponds to  
$\Lambda_{\rm QCD}$.
In this case, the boundary conditions are adopted such that
$\sigma(z) |_{z_0} = \partial_z \sigma(z) |_{z_m} = 0$, where $z_0$
is the UV cutoff which is taken to zero after all. Then the mass of
$\sigma$ depends on the parameters, $\lambda$ and $z_m$ since
$M^2_{\Phi}$ is fixed. The numerical results are shown in the 
Fig.\ref{vector}(B), where the spectrum of vector fields are also shown
in (A).
{In order to reproduce the experimental value 
$m_\sigma = (400-1200)$ MeV, $\lambda$ should be smaller than 0.35.}
The upper-bound for $m_\sigma$ is shown by the dotted line in the 
Fig.\ref{vector}(B).


\subsection{Vector meson}
As for the gauge bosons, 
they are separated to
the vector and the axial vector bosons $V_M$ and $A_M$, and are defined as 
$L_M \equiv V_M + A_M$ and $R_M \equiv V_M - A_M,$ respectively. 

\vspace{.3cm}
First of all, we discuss some properties of the vector mesons. 
The linearlized equation of motion for $V_{\mu}$ is (employing
$V_{z}=0$ gauge)
\beq
[(m^V_n)^2 + A^{-1}\partial_z A \partial_z]f_n^V=0,
\label{vectoreom}
\eeq
where the mode expansion $V_\mu(x,z)=\sum_{n}V_\mu^{(n)}(x) f_n^V(z)$ 
is applied and $m_n^V$ is the four dimensional mass of 
the $n$-th excited vector meson\footnote{
$f^V_n(z)$ is the wave-functions for vector mesons 
with the normalization condition
$1 = \int^{z_m}_{z_0}dz A(z)(f^V_n(z))^2$. }.
The boundary condition for vector mesons is given by
$f^V_n(z_0) = \partial_z f^V_n(z_m) = 0$, similarly for the
case of the sigma meson. 
Note that Eq.(\ref{vectoreom}) is independent of $v(z)$. Then
the masses of vector mesons
are given as the function of $\lambda$ and $z_m$ as in the case
of the $\sigma$ meson.

\vspace{.3cm}
It is convenient to solve (\ref{vectoreom}) numerically
with the above boundary conditions. The first excited vector meson mass,
$m_{\rho}$, is obtained as the function of $\lambda$ and $z_m$.
By using the experimental value of $m_{\rho}$, the IR cutoff $z_m$
is expressed as the function of $\lambda$. As the result, the second
and all highly excited vector meson masses depend on $\lambda$. 
We find that the lowest and the next excited $m_{\rho}$-meson are
fitted by the parameters $z_m=4$ and $\lambda=0.8$.

The vector meson decay constant is computed through the
second derivative of its own wave function according to \cite{EKSS}
\beq
F^2_{V_n} = \frac{1}{g_5^2} \left[ \left. 
\frac{d^2 f^V_n}{dz^2} \right|_{z_0} \right]^2, \qquad 
g_5^2 =\frac{12\pi^2}{N_c}. 
\eeq
Here $N_c =3$. However, its numerical value
given for $z_m=4$ and $\lambda=0.8$ deviates from the experimental value.
So we must change the value of $\lambda$ to the smaller side in order
to obtain more reliable result.

\begin{figure}[htbp]
\begin{center}
\voffset=15cm
  \includegraphics[width=6cm,height=5cm]{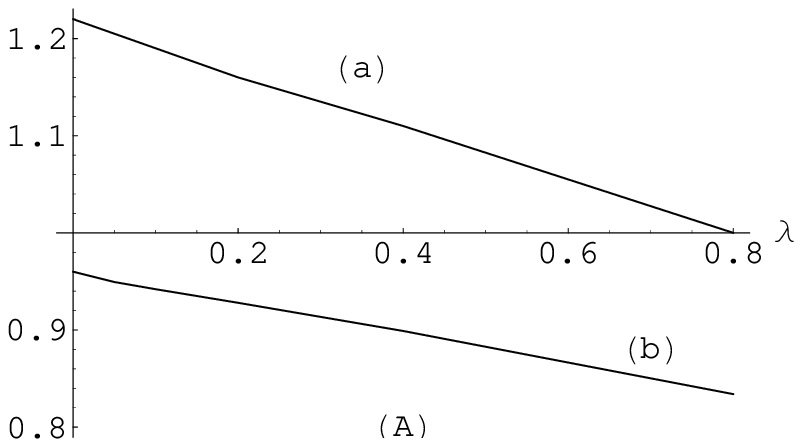}
  \includegraphics[width=6cm,height=5cm]{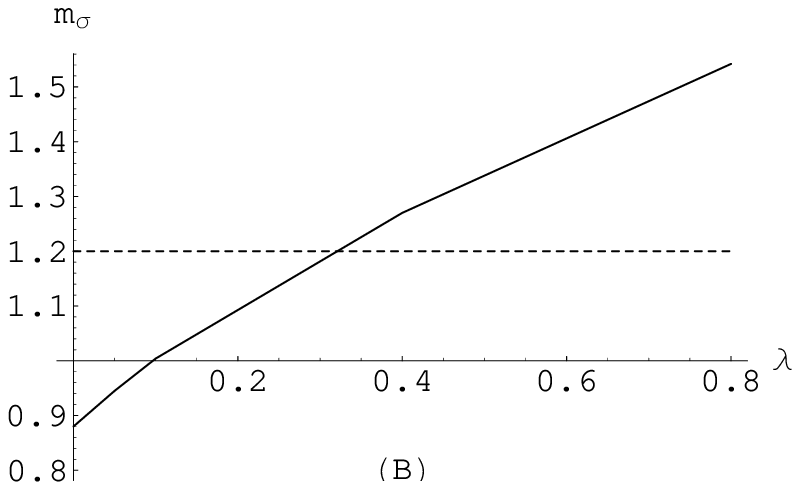}
\caption{(A) $\lambda$ dependences of the second excited vector meson 
$m_{V_2}$ (a) and the $\rho$ meson decay constant 
$F_\rho^{1/2}$ (b) 
divided by their experimental values~\cite{exp}, which are used
hereafter. 
(B) The curve and the dotted line
denote $m_{\sigma}$ (MeV) and its experimental upper bound, respectively.
}
\label{vector}
\end{center}
\end{figure}

Our results are shown in Fig. \ref{vector}, where we find 
that a good fit would be obtained near $\lambda \sim 0.8$ if we consider 
only the vector meson sector. {Note here that
in this sector, $v(z)$ (or equivalently,
the quark mass $m_q$ and/or the chiral condensate $c$) 
does not appear.
This is contrasted with the results from, for instance, 
the QCD sum rules \cite{KPT, RRY}, where the vector meson masses 
depend on the chiral condensate. This is due to the fact that the
spontaneous chiral symmetry breaking is introduced by hand in this
model, which was done by the choice of the profile of the scalar
field $S$.} 
In this sense, the analysis can not be complete at this stage.

\subsection{Axial-vector meson and $\pi$-meson}

The linearized equations of motion 
for the axial vector meson $A_\mu$ and the pion $\pi$ are obtained 
from the following quadratic action 
\beq
 S_{\rm axial}=\int dx^4dz\left[-{A \over 4g_5^2}(F_A^a)^2-
         {v^2 A^3\over 2}(\partial\pi^a+A^a)^2\right], \label{trancated-a}
\eeq
where $v(z)$ is given by (\ref{3-2}). 
Decomposing $A_\mu$ into the transverse and 
the longitudinal part, $A_\mu = A_{\mu \perp} + \partial_\mu \varphi$,
one can obtain equations for these fields, 
\bea
\label{axial1}
&&[
m_{A}^2+ A^{-1} \partial_z A \partial_z 
- g_5^2 A^2 v^2] A_{\mu \perp} = 0, \\
&&\partial_z (A \partial_z \varphi) - g_5^2 A^3 v^2 (\pi + \varphi) = 0, \\
\label{axial3}
&&
m_{\pi}^2\partial_z \varphi+ {g_5^2 A^2 v^2}\partial_z \pi=0. 
\eea
These equations are solved numerically under the boundary conditions, 
$A_{\mu \perp}(z_0)=\partial_zA_{\mu \perp}(z_m)=0,
~\varphi(z_0)=\partial_z\varphi(z_m)=\pi(z_0)=0$. 
The decay constants of the axial mesons and the pion are 
calculated from the wave functions as~\cite{EKSS} 
\beq
F_{A_n}^2 = \frac{1}{g_5^2} \left[ 
\left. \frac{d^2f^A_n}{dz^2} \right|_{z_0} \right]^2~(n \ne 0), \quad
F_\pi^2 = - \frac{1}{g_5^2} 
\left. \frac{\partial_z f}{z} \right|_{z_0} ,
\eeq
where $A_{\mu \perp}(x,z) = \sum_n A_\mu(x) f_n^A(z)$ and 
$f_n^A(z)$ is normalized as 
$1 = \int^{z_m}_{z_0}dz A(z) (f^A_n(z))^2$, and 
$f$ is the solution to Eq. (\ref{axial1}) with $m_A^2=0$, satisfying 
$f=0$ at $z=z_0$ and $df/dz=0$ at $z=z_m$.


The masses and the decay constants of axial-vector mesons and pion 
depend on four parameters 
$m_q$, $c$, $z_m$ and $\lambda$ through $v(z)$ and $A(z)$ 
in Eqs. (\ref{axial1})-(\ref{axial3}), 
while those of vector mesons do only on $z_m$ and $\lambda$. 
For the consistency between the vector and axial-vector meson sectors, here 
we take the same $z_m$ as that determined in the vector meson sector. 
The $z_m$ depends on $\lambda$ as 
$1/z_m=0.323-0.09125\lambda$~[GeV] for $\lambda < 0.8$. 

\begin{figure}[htbp]
\begin{center}
\voffset=15cm
  \includegraphics[width=8cm,height=5cm]{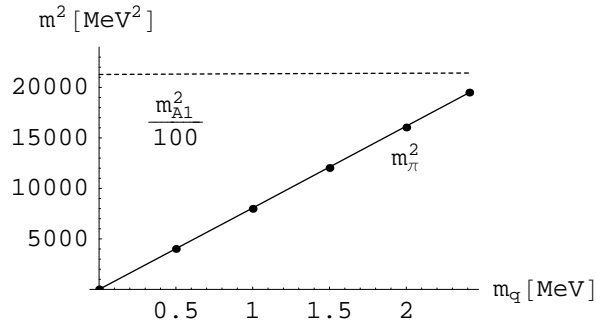}
\caption{The $m_q$ dependence of $m_\pi^2$ and $m_{A1}^2$. 
The solid line represents $m_{\pi}^2$, and 
the dotted line corresponds to $m_{A1}^2/100$. 
The GOR relation is shown by closed circles.
}
\label{GOR}
\end{center}
\end{figure}

First, we show that the present model reproduces 
the Gell-Mann-Oakes-Renner (GOR) relation, 
$m_\pi^2 F_\pi^2 = 2m_q c$, well satisfied in real QCD.
Figure \ref{GOR} shows the $m_q$ dependence of 
$m_\pi^2$, where other parameters are fixed 
at $\lambda=0.1$, $c=(0.3256~\rm {GeV})^3$
and $1/z_m = 0.315$~GeV. 
The solid line is a result of direct calculations of 
$m_\pi^2$, and the closed circles are a result obtained from 
calculated $F_\pi$ through the GOR relation $m_\pi^2 = 2m_q c/F_\pi^2 $. 
The two results agree with each other. Thus, the GOR relation is 
satisfied for the case of $\lambda=0.1$. 
This is also true for other $\lambda$, as shown later. 
In this figure, when $m_q=2.41$~MeV, 
calculated $m_\pi$ and $F_\pi$ reproduce the corresponding experimental 
values simultaneously. For comparison, we also plot the $m_q$ dependence of 
$m_{A1}^2$ by the dotted line. The $m_q$ dependence is quite weak. 
This means that the value of $m_{A1}$ is determined not by $m_q$ but by the 
chiral condensate $c$.

\begin{figure}[htbp]
\begin{center}
\voffset=15cm  
  \includegraphics[width=8cm,height=5cm]{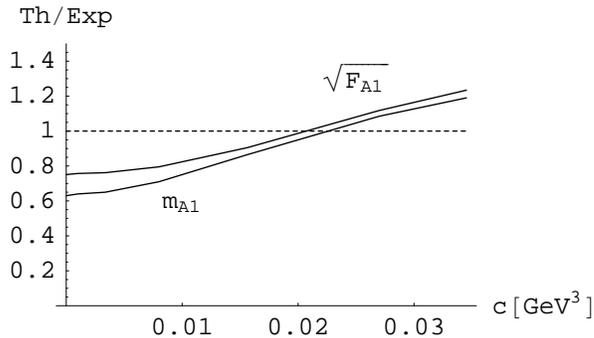}
\caption{The $c$ dependence of $F_{A_1}^{1/2}$ and $m_{A_1}$. 
These quantities are devided by the corresponding experimental 
values. 
}
\label{c-dep}
\end{center}
\end{figure}

Figure \ref{c-dep} shows the $c$ dependence of 
$m_{A_1}$ and $F_{A_1}^{1/2}$, in which other parameters are fixed 
at $\lambda=0.1$, $m_q=2.41~\rm {MeV}$
and $1/z_m = 0.315$~GeV.
The two quantities in the axial-vector sector have a similar $c$ dependence. 
They monotonously increase as $c$ increases, 
and they almost reproduce the corresponding experimental values at 
$c \sim 0.025~\rm{GeV}^3$. 
Mass $m_{A_1}$ is reduced to about 60 \% of the experimental value 
in the limit of 
no chiral condensate. Thus, 40 \% of the observed mass is generated by 
the finite chiral condensate.

We focus our analysis on the $\lambda$ dependence of $F_{A_1}, m_{A_1}, F_\pi$. 
Parameters $m_q$ and $c$ are fixed as follows. 
First, $c$ is assumed to be determined from $m_q$ through 
GOR relation, 
$\bar{m}_\pi^2 \bar{F}_\pi^2 = 2m_q c$, 
where $\bar{m}_\pi$ and $\bar{F}_\pi$ denote experimental values of 
$m_\pi$ and $F_\pi$. 
Second, $m_q$ is fixed so as to 
reproduce the observed pion mass. The resultant $m_q$ depends on 
$\lambda$ as $m_q=2.26+1.125\lambda$~[MeV] for $\lambda < 0.4$.

\begin{figure}[htbp]
\begin{center}
\voffset=15cm
  \includegraphics[width=7cm,height=5cm]{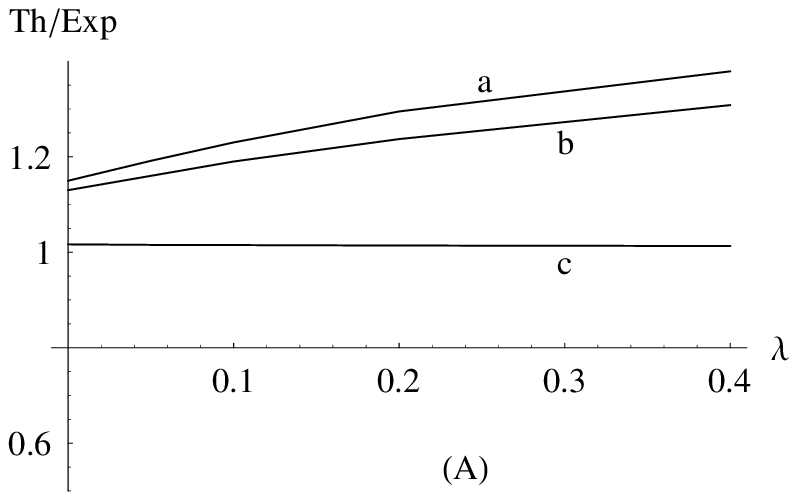}
  \includegraphics[width=7cm,height=5cm]{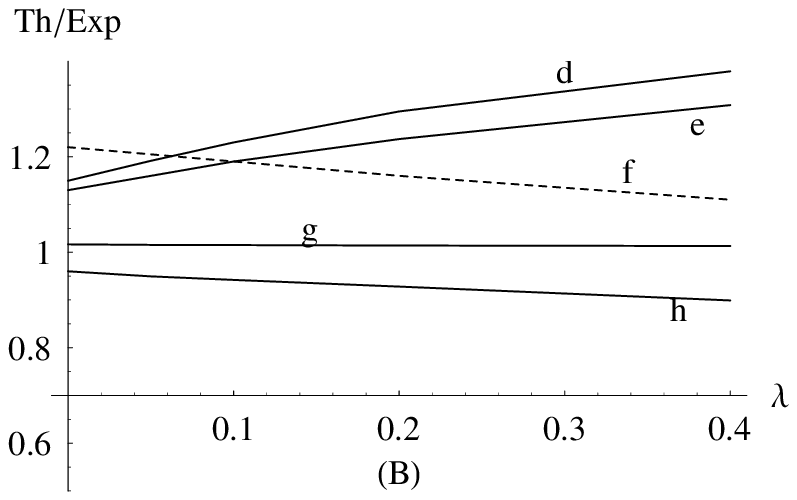}
\caption{Predictions of the model in the axial-vector sector (A)
and in both the vector and axial-vector meson sectors (B). 
As for (A), the solid lines, a, b, c, denote $F_{A_1}^{1/2}$, $m_{A_1}$, 
$F_\pi$, respectively. 
As for (B), the solid lines, d, e, g, h, represent 
$F_{A_1}^{1/2}$, $m_{A_1}$, $F_\pi$, $F_\rho^{1/2}$, 
respectively, and the dotted line, f, corresponds to $m_{V_2}$. 
All quantities are divided by the corresponding experimental 
values.
}
\label{axialall}
\end{center}
\end{figure}

Figure \ref{axialall}(A) shows $\lambda$ dependence of 
predicted values of $F_{A_1}^{1/2}, m_{A_1}, F_\pi$. 
As for all $\lambda$ up to 0.4, 
calculated $F_\pi$ agrees with 
the experimental value with $\sim 1 \%$ error. 
In the present analysis, we assumed the calculated $F_\pi$ to 
reproduce the experimental value when we used the GOR relation as a relation 
between $c$ and $m_q$. 
The good agreement shows that our assumption is consistent and 
then the GOR relation is well satisfied for all $\lambda$ at least up to 0.4. 
As for $F_{A_1}^{1/2}$ and $m_{A_1}$, the agreement of the theoretical results with the observed ones becomes better as $\lambda$ decreases. 
One then see that the $\lambda=0$ case, namely the AdS${}_5$ case, 
yields a best fit if we consider only the axial-vector meson sector.

Figure \ref{axialall}(B) summarizes all predictions of our model 
in both the vector and axial-vector meson sectors.  
As for three of five quantities, $F_{A_1}^{1/2}, m_{A_1}, F_\rho^{1/2}$, 
the agreement of the predictions with the corresponding observations 
is best at $\lambda=0$, but as for one of the five, i.e. for $m_{V_2}$ 
the best fit is obtained at $\lambda \sim 0.8$. 
Hence, we can conclude that totally the best fit 
is realized at small $\lambda$. 
This result is welcomed to obtain a realistic 
mass of $\sigma$ meson shown in Fig \ref{vector}.

\section{Summary}

We present a new holographic model of QCD in 
which the gravity-background is deformed from pure AdS$_{5}$ 
and the extra dimension is cut off at an appropriate infrared point. 
This is a natural extension of 
the holographic models of Refs. \cite{EKSS,RP,TB,HS} 
in which the AdS$_{5}$ background is taken and the extra dimension is 
cut off in the same manner. 
These holographic models based on AdS$_{5}$ 
well reproduced observed quantities on light mesons. Nevertheless, 
there is no strong reason why AdS$_{5}$ is taken as a background of 
the gravity dual to QCD, 
since QCD has a running coupling constant and then the gravity-background 
should be modified from AdS$_{5}$.  
Our new model is proposed to answer this question.

In our model, parameter $\lambda$ represents the magnitude of 
the deformation of AdS$_{5}$ background. 
For any $\lambda$, 
our model reproduces the GOR relation which is well satisfied in 
real QCD. 
As for light mesons except the excited $\rho$ meson, 
a best fit of our predictions to 
the corresponding observed quantities is obtained 
at $\lambda=0$, that is, in the AdS$_{5}$ limit. 
Meanwhile, the excited $\rho$ meson mass is reproduced at $\lambda=0.8$. 
Hence, a best fit to all of them is realized at small $\lambda$. 
The parameter $\lambda$ is related to 
the five-dimensional Planck mass $M_5$ and the scale 
of the supersymmetry breaking $\Lambda_{\rm SUSY}$ 
as $\lambda=\Lambda_{\rm SUSY}^5/(3 M_5^3)$.
The smallness of $\lambda$ would imply $M_5 \gg \Lambda_{\rm SUSY}$ 
in the gravity dual to QCD.

It is well known that the chiral condensate $c$ plays 
an important role on all quantities of light mesons. In this sense, 
the quantities should depend on $c$. 
Our model has the property for the axial-vector meson sector, 
but not for the vector meson sector. Thus, the present model 
is insufficient for the vector meson sector. 
The holographic models of Refs. \cite{EKSS,RP,TB,HS} have 
essentially the same problem. This is an important problem to 
be solved in future.

\vspace{.3cm}
\section*{Acknowledgments}
{M. T. would like to thank T. Hatsuda and D. T. Son for valuable
information.} This work has been supported in part by the Grants-in-Aid
for Scientific Research (13135223)
of the Ministry of Education, Science, Sports, and Culture of Japan.
N.M. was supported by RIKEN (No.A12-61014).



\begin{thebibliography}{99}

\bibitem{KK}
  A.~Karch and E.~Katz, 
  JHEP {\bf 0206}, 043(2003) [hep-th/0205236].

\bibitem{KMMW}
  M.~Kruczenski, D.~Mateos, R.C.~Myers and D.J.~Winters, 
  JHEP {\bf 0307}, 049(2003) [hep-th/0304032].

\bibitem{KMMW2}
  M.~Kruczenski, D.~Mateos, R.~C.~Myers and D.~J.~Winters,
  JHEP {\bf 0405}, 041 (2004) [arXiv:hep-th/0311270].

\bibitem{Bab}
  J.~Babington, J.~Erdmenger, N.~J.~Evans, Z.~Guralnik and I.~Kirsch,
  Phys.\ Rev.\ D {\bf 69}, 066007 (2004) [arXiv:hep-th/0306018].

\bibitem{ES}
  N.~J.~Evans and J.~P.~Shock,
  Phys.\ Rev.\ D {\bf 70}, 046002 (2004) [arXiv:hep-th/0403279].


\bibitem{NPR}
  C.~Nunez, A.~Paredes and A.V.~Ramallo, 
  JHEP {\bf 0312}, 024(2003) [hep-th/0311201]. 
  
\bibitem{SSu}
 T. Sakai and S. Sugimoto, Prog.Theor.Phys.113(2005)843-882,
 [hep-th/0412141];  hep-th/0507073. 

\bibitem{GY2} 
  K. Ghoroku and M. Yahiro, 
  Phys.\ Lett.\ B {\bf 604}, 235 (2004) [arXiv:hep-th/0408040].

\bibitem{EKSS}
  J.~Erlich, E.~Katz, D.~T.~Son and M.~A.~Stephanov,
   hep-ph/0501128.

\bibitem{RP}
  L.~Da Rold and A.~Pomarol,
   Nucl.\ Phys.\ B {\bf 721}, 79 (2005) [hep-ph/0501218];
   hep-ph/0510268.
\bibitem{TB}
  G.~F.~de Teramond and S.~J.~Brodsky,
  Phys.\ Rev.\ Lett.\  {\bf 94}, 201601 (2005) [arXiv:hep-th/0501022].

\bibitem{HS}
  J.~Hirn and V.~Sanz, 
  hep-ph/0507049.


\bibitem{GTU}
  K.~Ghoroku, M.~Tachibana and N.~Uekusa,
   Phys.\ Rev.\ D {\bf 68}, 125002 (2003) [hep-th/0304051].

\bibitem{gppz1}
  L. Girardello, M. Petrini, M. Porrati and A. Zaffaroni,  
  JHEP 9812 (1998) 022 [hep-th/9810126]; 
  JHEP 9905 (1999) 026 [hep-th/9903026]. 

  
%
%



\bibitem{GY}
  K.~Ghoroku and M.~Yahiro,
   Class.\ Quant.\ Grav.\ {\bf 20}, 3717 (2003) [hep-th/0211112].

\bibitem{KPT}
  N. V. Krasnikov, A. A. Pivovarov and N. N. Tavkhelidze, 
  Z.\ Phys.\ C19, 301 (1983). 

\bibitem{RRY}
  L. J. Reinders, H. Rubinstein and S. Yazaki,  
  Phys.\ Rept.\ 127, 1 (1985). 
\bibitem{exp}
S. Eidelman et al. [Particle Data Group Collaboration],
Phys. Lett. {\bf B592}, 1 (2004). \\





\end{thebibliography}
\end{document}